\documentclass[conference]{IEEEtran}
\IEEEoverridecommandlockouts
\usepackage{cite}
\usepackage{amsmath,amssymb,amsfonts}
\usepackage{algorithmic}
\usepackage{graphicx}
\usepackage{textcomp}
\usepackage{xcolor}
\usepackage{tikz}
\usepackage{pgfplots}
\usetikzlibrary{arrows}
\usetikzlibrary{datavisualization.formats.functions}
\pgfplotsset{compat=1.5}
\def\BibTeX{{\rm B\kern-.05em{\sc i\kern-.025em b}\kern-.08em
    T\kern-.1667em\lower.7ex\hbox{E}\kern-.125emX}}
\begin{document}

\title{A Weighted Autoencoder-Based Approach 
to Downlink NOMA Constellation Design 

\thanks{This article/publication is based upon work from COST Action INTERACT, CA20120, supported by COST (European Cooperation in Science and Technology).\\
This paper has received funding from the European Union’s Horizon 2020 research and innovation programme under Grant Agreement number 856967.\\
This work was supported by Grant PID2021-128373OB-I00 (6G-AINA) funded by
MCIN/AEI/10.13039/501100011033 and by “ERDF A way of making Europe,” }
}

\author{\IEEEauthorblockN{Vukan Ninkovic, Dejan Vukobratovic}
\IEEEauthorblockA{\textit{Faculty of Technical Sciences} \\
\textit{University of Novi Sad}\\
Novi Sad, Serbia \\
\{ninkovic, dejanv\}@uns.ac.rs}
\and
\IEEEauthorblockN{Adriano Pastore, Carles Ant\'on-Haro}
\IEEEauthorblockA{
\textit{Centre Tecnol\`{o}gic de Telecomunicacions de Catalunya (CTTC/iCERCA)}\\
\textit{Parc Mediterrani de la Tecnologia (PMT), Av. Carl Friedrich Gauss 7, Building B4}\\
08860 Castelldefels, Spain \\
\{adriano.pastore, carles.anton\}@cttc.es}
}

\maketitle

\begin{abstract}
End-to-end design of communication systems using deep autoencoders (AEs) is gaining attention due to its flexibility and excellent performance. Besides single-user transmission, AE-based design is recently explored in multi-user setup, e.g., for designing constellations for non-orthogonal multiple access (NOMA). In this paper, we further advance the design of AE-based downlink NOMA by introducing weighted loss function in the AE training. By changing the weight coefficients, one can flexibly tune the constellation design to balance error probability of different users, without relying on explicit information about their channel quality. Combined with the SICNet decoder, we demonstrate a significant improvement in achievable levels and flexible control of error probability of different users using the proposed weighted AE-based framework.
\end{abstract}


\section{Introduction}

Improving the resource usage efficiency of 5G and beyond-5G wireless systems is essential to accommodate increasing user requirements. Non-orthogonal multiple access (NOMA) is a recent addition to the wireless signal processing toolset that promises further advances in achieving improved spectrum usage under additional benefits of reduced latency, increased throughput, higher connection density and improved fairness \cite{Liu, Ding}. The main idea behind NOMA is that multiple users can be served by the same resources if appropriate joint signal encoding (e.g., using superposition encoding principle) and efficient signal decoding (e.g., using successive interference cancellation) is applied \cite{NOMA_Survey,Vanka}.

Recent trends see shifting the design of encoding and decoding procedures from conventional to machine learning (ML)-based methods. The trend is initiated in the domain of point-to-point communication systems \cite{OShea_2017,Doerner}, but has since expanded to multi-user NOMA setup \cite{Van Luong, Alberge,Kim,Gui}. In this paper, we follow this trend, and use a deep autoencoder (AE)-based approach to design encoding and decoding solution for downlink NOMA. Building upon the work in \cite{Alberge}, we apply: 1) a weighted loss function to control error probability balance across different users, 2) SICNet architecture \cite{Van Luong} to enhance deep AE-based decoding capability. Using the proposed weighted AE approach, we are able to obtain significant improvement and flexibility in the error rate performance, as evidenced by simulation experiments.

The paper is organised as follows. In Sec. II, we provide background on generic NOMA system, and recent ML-based methods we use in our work. In Sec. III, we represent NOMA system as end-to-end AE-based system and introduce our weighted AE approach combined with SICNet. Performance evaluation and comparison with the baseline AE-based NOMA is presented in Sec. IV. The paper is concluded in Sec. V.

\section{Background and System Model}

\subsection{Downlink NOMA Transmission}
\label{NOMA DL}

    We consider a problem of downlink NOMA transmission of $L$ (different) messages from a base station (BS) to $L$ users. The BS jointly encodes user messages into a signal $\mathbf{x}$ and transmits it over $n$ channel uses. We assume a user message $\mathbf{s}_{\ell} \in \mathbb{F}_2^{k_{\ell}}, \ell \in \{1,2,\ldots,L\}$ is a binary (information) sequence of length $k_{\ell}$ bits. 
    
    Encoding process for the 
    $\ell^{\textrm{th}}$ user can be described as 
    $f_{\ell}:\mathbb{F}_2^{k_{\ell}} \rightarrow \mathbb{R}^n$. In other words, the $\ell$-th user maps its information sequence $\mathbf{s}_{\ell}$ into a signal $\mathbf{q}_{\ell}$ obtained as $\mathbf{q}_{\ell}=f_{\ell}(\mathbf{s}_{\ell}), \mathbf{q}_{\ell}\in \mathbb{R}^n, \ell \in \{1,2,\ldots,L\}$. Based on the $L$ signals ($\mathbf{q}_1, \mathbf{q}_2, \ldots, \mathbf{q}_{\textrm{L}}$), the BS generates the transmitted message $\mathbf{x}$ of length $n$ ($\mathbf{x}\in\mathbb{R}^{n}$), usually by exploiting generic superposition coding function $F(\cdot)$:

\begin{align}
\label{F_ref}
    \mathbf{x}=F(f_1(\mathbf{s}_1), \ldots, f_{\textrm{L}}(\mathbf{s}_{\textrm{L}}))=F(\mathbf{q}_1, \ldots, \mathbf{q}_{\textrm{L}}),
\end{align}
where in the case of conventional superposition coding the function $F(\cdot)$ represents a weighted linear combination \cite{Vanka}. At the output of the encoder, $\mathbf{x}$ obeys an average power constraint, for which, $\frac{1}{M_c}\sum_{i=1}^{i=M_c} \|\mathbf{x}_i \|_2^2 = n$, where $M_c=2^{k_1+k_2+\ldots+k_{\textrm{L}}}$. 

Each of the $L$ users has distinct channel conditions, which transforms input message $\mathbf{x} \in\mathbb{R}^{n}$ into $L$ output sequences $\mathbf{y}_{\ell} \in\mathbb{R}^{n}, \ell \in   \{1,2,\ldots,L\}$ following the probabilistic channel model $p(\mathbf{y}_{\ell}|\mathbf{x})$. In this paper, we consider a set of independent additive white Gaussian noise (AWGN) channels between the BS and each of the $L$ users characterised by the set of noise variances $\{\sigma^2_\ell\}_{\ell=1,\ldots,L}$ \cite{Shieh}. This model is suitable for downlink NOMA in OFDM-based systems, and it can be extended to a set of independent block-fading channels \cite{Alberge}.

At the receiver side, an estimated information sequence for the $\ell^{\textrm{th}}$  user is obtained via the function $g_{\ell}:\mathbb{R}^n \rightarrow \hat{\mathbf{s}}_{\ell}$, i.e., the decoding process for each user can be defined as $\hat{\mathbf{s}}_{\ell}=g_{\ell}(\mathbf{y}_{\ell}), \ell \in \{1,2,\ldots,L\}$. 

The goal is to design a combination of encoders $\{\{f_\ell\}_{\ell=1,\ldots,L},F\}$ and decoders $\{g_\ell\}_{\ell=1,\ldots,L}$ that will optimally balance (in a Pareto optimal sense) the set of message error probabilities $\{P_{\textrm{e}_{\ell}}\}_{\ell=1,\ldots,L}$:

\begin{align}
\label{messErr}
    P_{\textrm{e}_{\ell}} = \frac{1}{2^{k_{\ell}}} \sum_{\mathbf{s}_{\ell} \in \mathbb{F}_2^{k_{\ell}}} \mathbb{P}\{\hat{\mathbf{s}}_{\ell} \neq \mathbf{s}_{\ell}|\mathbf{s}_{\ell}\}
\end{align}
across the set of $L$ receivers.

\subsection{Power--Domain Downlink NOMA}
\label{powNOMA}

     
In the conventional downlink NOMA scheme, the function $F(\cdot)$ defined in Eq. \ref{F_ref} is a weighted linear combination of user signals obtained by allocating different power coefficients as the weights of the linear combination \cite{NOMA_Survey,Ding}:

    \begin{align}
    \label{eq1}
    \mathbf{x}=F(\mathbf{q}_1, \ldots, \mathbf{q}_{\textrm{L}})=\sum_{\ell=1}^{\ell=L}\sqrt{p_{\ell}}\mathbf{q}_{\ell},
    \end{align}
    where $p_{\ell}$ represents power associated with user $\ell$.
    Without loss of generality, we assume that the power is allocated to users in ascending order $p_1<p_2<p_3<\ldots<p_{\textrm{L}}$ \cite{Liu}.
    From the receiver perspective, the $\ell^{\textrm{th}}$  user observes:
    \begin{align}
    \label{eq2}\mathbf{y}_{\ell}=h_{\ell}\mathbf{x}+\mathbf{z}=h_{\ell}\sum_{\ell=1}^{\ell=L}\sqrt{p_{\ell}}\mathbf{q}_{\ell}+\mathbf{z}
    \end{align}
    where $h_{\ell}$ represents complex channel coefficient between BS and user $\ell$, while additive white Gaussian noise (AWGN) with variance $\sigma^2_{\ell}$ is denoted as $\mathbf{z}$. For simplicity and without loss of generality, we suppose that $h_{\ell}=1, \ell \in \{1, 2, \ldots, L\}$, i.e., we focus on the AWGN channel (Section \ref{NOMA DL}).
 
 A combination of superposition coding at the BS and successive interference cancellation (SIC) decoding at the receiver is commonly used in downlink NOMA \cite{Liu}. SIC exploits different powers allocated to different users at the transmitter side. The user with the power coefficient $p_{\textrm{L}}$ (the strongest power coefficient) will be decoded first (treating other users signals as noise) and subtracted from $\mathbf{x}$. This procedure is successively repeated for the remaining users until all user messages are decoded \cite{NOMA_Survey,Vanka}. 
Note that the SIC performance is dependent on a perfect knowledge of channel state information (CSI) and power allocation coefficients, and may degrade if imperfect CSI is used \cite{Yang}. 

\subsection{SICNet}
\label{SIC}

In order to overcome the above-mentioned problem of the conventional SIC receiver, the authors in \cite{Van Luong} proposed a deep neural network (DNN)-based decoding algorithm, entitled \textit{SICNet}, that 
incorporates DNNs to recover user messages. \textit{SICNet} preserves original SIC decoder structure and algorithmic flow, however, each block is replaced with a DNN that performs classification task \cite{Van Luong}. 
Output of each DNN can be used as a soft estimate (by using softmax activation function) and concatenated with the input of the subsequent DNN. In such a way, the input to the \textit{SICNet} of $\ell^{\textrm{th}}$ user is its received signal $\mathbf{y}_{\ell}$ concatenated with soft estimates of users $\ell+1$ to $L$. This approach avoids the explicit need for prior CSI and power allocation knowledge \cite{Van Luong}.

\subsection{Autoencoder--Based Communication Systems}
\label{AEComm}

\begin{figure}[t]
	\centering
	\includegraphics[width=1\linewidth]{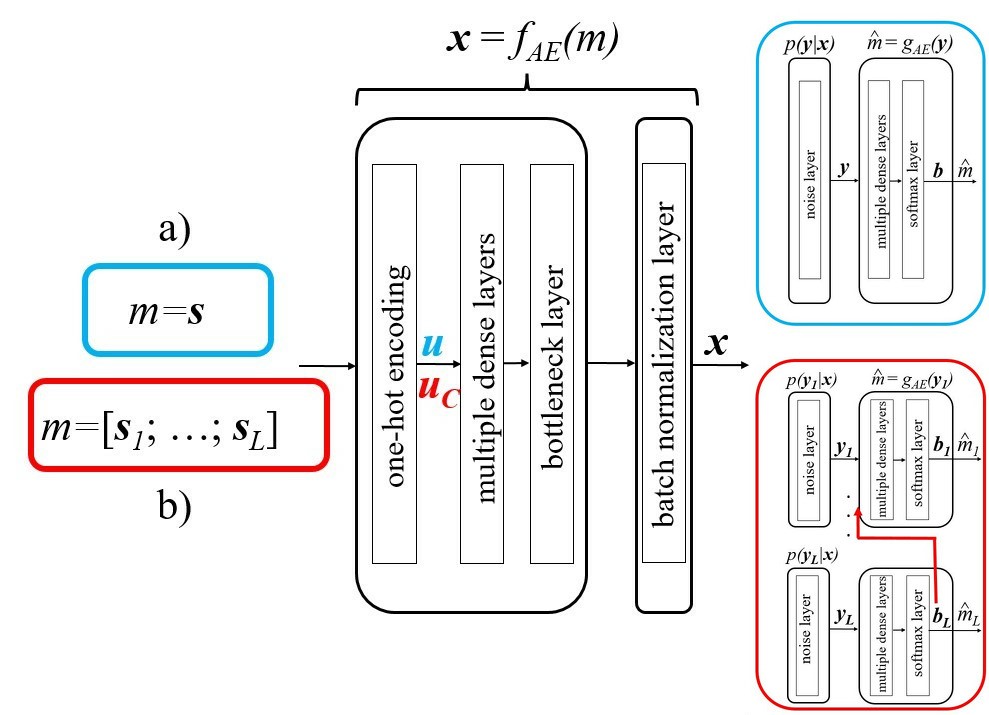}
	\caption{Communication system represented as a deep autoencoder: a) Conventional system - Blue blocks \cite{OShea_2017} b) NOMA downlink system with \textit{SICNet} \cite{Van Luong} - Red blocks; Black blocks are used by both a) and b)}
	\label{Fig_AE}
\end{figure}

The conventional point-to-point communication system with a transmitter and a receiver has been redesigned from the deep learning perspective in the form of an end-to-end autoencoder (AE) \cite{OShea_2017}, as illustrated in Fig. \ref{Fig_AE}a (blue blocks). 

An input message $m$, which can be represented as a sequence of $k$ bits $\mathbf{s}=(s_1,s_2,\ldots,s_\textrm{k})$ (where $k=\log_2{M}$), is encoded as a one-hot vector $\mathbf{u} = (u_1,u_2,\ldots,u_\textrm{M}) \in \{0,1\}^M$. The transmitter, described by $f_{\textrm{AE}}:m \rightarrow \mathbf{x}$, is represented as a feed-forward neural network with $H$ hidden layers, followed by a bottleneck layer of width $n$. At the output of transmitter, a batch normalization layer ensures that the average power constraint on $\mathbf{x}$ are met (Section \ref{NOMA DL}).
    
    The AWGN channel is constructed as an additive noise layer with output $\mathbf{y}=\mathbf{x}+\mathbf{z}$ (as it is shown in Fig. \ref{Fig_AE}, blue receiver), where $\mathbf{z}$ contains $n$ independent and identically distributed  samples of a Gaussian random variable with zero mean and variance $\sigma^2$. 
    
    The receiver, described as $g_{\textrm{AE}}:\mathbf{y} \rightarrow \hat{m}$, is implemented as a feed-forward neural network with $H$ hidden layers and softmax activation function at its output, $\mathbf{b} = (b_1,b_2,\ldots,b_\textrm{M}) \in (0,1)^M$, $\Vert \mathbf{b}\Vert_1=1$. All other hyperparameters are the same as in the AE transmitter part. The receiver takes the channel output $\mathbf{y}$ as its input and produces a message estimate $\hat{m}$ as $\hat{m}=\arg\max_{i}\{b_i\}$.
    
    Except the output layer of the transmitter and receiver that have linear and softmax activation function, respectively, all other layers are activated using rectified linear units (ReLU).

    The presented AE is trained in an end-to-end manner  to minimize the message error probability $P_{\textrm{e}}$ (Eq. \ref{messErr}), by using stochastic gradient descent with Adam optimizer \cite{adam}. In other words, $f_{\textrm{AE}}$ and $g_{\textrm{AE}}$ are jointly optimized. In order to optimize $P_{\textrm{e}}$ (and parameters of the AE), the minimization of categorical--cross entropy between $\mathbf{u}$ and $\mathbf{b}$ is used as a surrogate for the message error probability \cite{OShea_2017}: 

\begin{align}\label{eq3}
\ell(\mathbf{u},\mathbf{b})=-\sum_{i=1}^{M} u_i \log{b_i},    
\end{align}

\section{Autoencoder--Based Downlink (DL) NOMA}

    In this section, we present a flexible and efficient way to learn encoding and decoding strategy for NOMA downlink transmission. Conventional NOMA downlink communication system with SIC decoding, described in Sections \ref{NOMA DL} and \ref{powNOMA}, can be implemented in a DNN fashion as an extension of the  end-to-end AE-based scheme presented in Section \ref{AEComm}, as illustrated in Fig. \ref{Fig_AE}b (red blocks)\cite{Alberge}. 
    
    Starting from the superposition coding at the transmitter side, a composition of the function $F(\cdot)$ and single-user encoding functions $\{f_{\ell}\}_{\ell=1, 2, \ldots, L}$ (Eq. \ref{F_ref}), is replaced with a single learning process. More precisely, an encoder is jointly optimized with the set of $L$ decoders using an end-to-end AE-based approach, i.e., the function pairs $(f_{\textrm{AE}}, \{g_{AE_\ell}\}_{\ell=1, 2, \ldots, L})$ are obtained using an end-to-end training procedure. Individual users messages $\mathbf{s}_{\ell}\in\mathbb{F}_2^{k_{\ell}} , \ell \in \{1,2,\ldots,L\},$ are represented as a message index from a set $\mathcal{M}_{\ell}, |\mathcal{M}_\ell|=M_{\ell}=2^{k_{\ell}},$ and one-hot encoded, where $\mathbf{u}_{\ell}$ denotes one-hot encoding form of the $\ell^{\textrm{th}}$  user message $\mathbf{s}_{\ell}$, and $\mathbf{U}=\{\mathbf{u}_1, \mathbf{u}_2, \ldots, \mathbf{u}_{\textrm{L}}\}$. AE input message $m$ is defined as a concatenation of individual user messages in their bit representations, i.e., $m=[\mathbf{s}_1;\mathbf{s}_2;\ldots;\mathbf{s}_{\textrm{L}}]$ (Fig. \ref{Fig_AE}b)  \cite{Alberge}.
    One-hot encoding of the AE message $m$ maps $k_C=k_1+k_2+\ldots+k_{\textrm{L}}$ bits into a one-hot encoding $\mathbf{u}_C= (u_1,u_2,\ldots,u_{M_C}) \in \{0,1\}^{M_C}$ where $M_C=2^{k_C}$.
    The rest of the transmitter is the same as in Section \ref{AEComm} (Fig. \ref{Fig_AE}, black blocks).
    
    The transmitted message $\mathbf{x}=f_{\textrm{AE}}(m)$ is passed through $L$ different AWGN channels whose outputs $\mathbf{y}_1, \mathbf{y}_2, \ldots, \mathbf{y}_{\textrm{L}}$ are available at the respective receivers. Without loss of generality, we suppose that each of the $L$ channels has different signal-to-noise ratio ($SNR$) and $SNR_1>SNR_2>\ldots>SNR_{\textrm{L}}$.  
    
    At the receiver side, each of the $L$ users estimates the corresponding message through the feed-forward neural network, described in Section \ref{AEComm} (there are in total $L$ DNN receivers, one per each user). 
    In order to implement \textit{SICNet} \cite{Van Luong} (Section \ref{SIC}), soft output of each preceding users' DNN is connected to the input of the next user (Fig. \ref{Fig_AE}-red block receiver), where the decoding process starts from the user with the most degraded channel ($\mathbf{y}_{\textrm{L}}$).
    The softmax output of all $L$ users is collected into $\mathbf{B}=\{\mathbf{b}_1, \mathbf{b}_2, \ldots, \mathbf{b}_{\textrm{L}}\}$, where each $\mathbf{b}_{\ell}$ is of length $M_{\ell}=2^{k_{\ell}}, \ell\in \{1,2,\ldots,L\}$.
    
    The goal is to learn a pair ($f_{\textrm{AE}}, \{g_{AE_\ell}\}_{\ell=1, 2, \ldots, L}$) that minimizes the individual user message error probabilities $\{P_{\textrm{e}_{\ell}}\}_{\ell=1, 2, \ldots, L}$ (Eq. \ref{messErr}). As in Section \ref{AEComm}, minimization of  $P_{\textrm{e}_{\ell}}$ is replaced with the loss function based on the categorical-cross entropy, defined for the $\ell^{\textrm{th}}$  user as:

\begin{align}\label{eq4}
\ell_{\ell}(\mathbf{u}_{\ell},\mathbf{b}_{\ell})=-\sum_{i=1}^{M_{\ell}} u_{{\ell_i}} \log{b_{{\ell_i}}},    
\end{align}

The total loss function for the AE-based downlink NOMA is the sum of the individual users' loss contributions:
    
\begin{align}
\label{loss}
    L(\mathbf{U}, \mathbf{B})=\sum_{j=1}^L \ell_{j}(\mathbf{u}_j,\mathbf{b}_j)
\end{align}

\subsection{Weighted Autoencoder-Based Design of Downlink NOMA}

    In order to train AE-based downlink NOMA communication system described in the previous section, the SNR difference $\Delta SNR$ between subsequent receivers is needed as one of the hyperparameters of the training process. As channel conditions may change during the testing phase, this may lead to a potential performance degradation.
    
    In our previous work \cite{Ninkovic_2021}, we explored a novel class of autoencoders that utilise a compound loss function in the context of single-user unequal error protection (UEP) coding. Such AE method is able to flexibly balance between error probabilities among the set of messages (message-wise UEP) or between specific subblocks of different importance classes within a single message (bit-wise UEP). Inspired with this approach, a similar weighted sum approach can be used in the AE-based downlink NOMA by introducing weights associated to $L$ different users (Eq. \ref{loss}), leading to a weighted total loss function:

\begin{align}
\label{loss_weight}
        L(\mathbf{U}, \mathbf{B})=\sum_{j=1}^L \lambda_j\ell_{j}(\mathbf{u}_j,\mathbf{b}_j),
\end{align}
where $\lambda_{j}$ represents a weight factor associated to the user $j$, $\sum_{j=1}^L \lambda_j = 1$, and $\lambda_j \geq 0$.

Loss function defined in Eq. \ref{loss} is a special case of the weighted sum presented in Eq. \ref{loss_weight}, where $\lambda_{\ell}=\frac{1}{L}, \ell\in\{1, 2, \ldots, L\}$. By incorporating the above mentioned compound loss function (Eq. \ref{loss_weight}), all users can be trained assuming they experience the same SNR (thus disregarding $\Delta SNR$ as a hyperparameter) while adjusting the desirable users' performance using $\mathbf{\lambda}=\{\lambda_1, \ldots, \lambda_{\textrm{L}}\}$. Although a simple alteration of loss function, the proposed approach introduces a single ``knob'' (parameter $\lambda$) one can tune to design a family of NOMA constellations that progressively balance the users' error probabilities. The influence of $\lambda$ on the system performance is elaborated in the next section.

\section{Performance Evaluation of the Weighted AE--Based Downlink NOMA}

\subsection{Training Procedure}

For simplicity, and without loss of generality, the number of users in all conducted experiments is restricted to two ($\mathbf{\lambda}=\{\lambda_1, \lambda_2\}=\{\lambda, 1-\lambda\}$), and $k_1=k_2=k$ (number of messages is $M_c=2^{k_1+k_2}=2^{2k}$). Apart from the introduction of the suitable loss function (Eq. \ref{loss} and \ref{loss_weight}), the same training procedure as presented in \cite{OShea_2017} is preserved, i.e., AE-based NOMA downlink system is optimized by using stohastic gradient descent with Adam optimizer\cite{adam} (learning rate $\alpha=0.0009$, $\beta_1=0.9$, $\beta_2=0.999$). Regarding the number of bits associated to the each user, for both encoder and each of two decoders, single fully--connected hidden layer is considered ($H=1$) with $M_c$ neurons. Batch size is set to 3000. When Eq. \ref{loss} is used as a loss function, training is performed at $\Delta SNR=SNR_1-SNR_2=9$ dB, where $SNR_1=15$ dB. With introduction of $\lambda$ (Eq. \ref{loss_weight}) both decoders are trained on the same $SNR_1=SNR_2=6$ dB. During the testing phase, $\Delta SNR$ is preserved at $9$ dB for all conducted experiments. 
Training and test data sets contain $10^5$ and $2\times10^6$ messages sampled at random from $\{1, 2, \ldots, M_c\}$, respectively. 

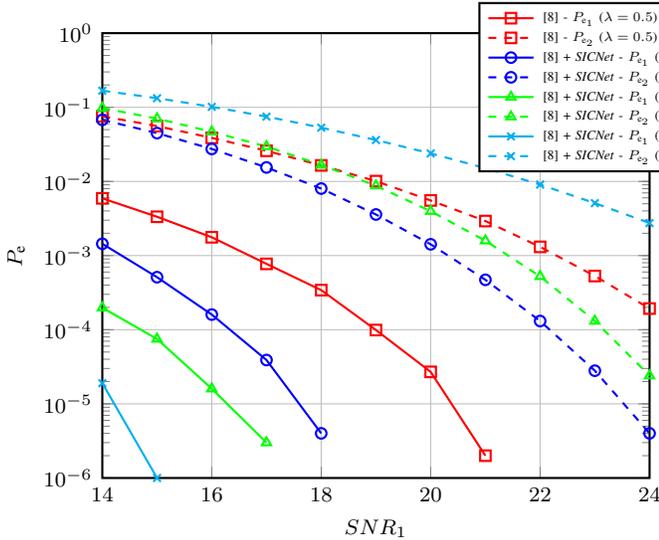
\begin{figure}[t]
	\begin{tikzpicture}
  	\begin{semilogyaxis}[width=1\columnwidth, height=7.5cm, 
	legend style={at={(0.92,1.07)}, anchor= north,font=\scriptsize, legend style={nodes={scale=0.72, transform shape}}},
   	legend cell align={left},
   	x tick label style={/pgf/number format/.cd,
   	set thousands separator={},fixed},
   	y tick label style={/pgf/number format/.cd,fixed, precision=2, /tikz/.cd},
   	xlabel={$SNR_1$},
   	ylabel={$P_{\textrm{e}}$},
   	label style={font=\footnotesize},
   	grid=major,   	
   	xmin = 14, xmax = 24,
   	ymin=0.000001, ymax=1,
   	line width=0.8pt,
   	tick label style={font=\footnotesize},]
   	\addplot[red, mark=square] 
   	table [x={x}, y={y}] {./Figs/n2_k2/bler1_fr.txt};
   	\addlegendentry{\cite{Alberge} - $P_{\textrm{e}_1}$         ($\lambda=0.5$)}
        
        \addplot[dashed,red, mark=square, mark options={solid}] 
   	table [x={x}, y={y}] {./Figs/n2_k2/bler2_fr.txt};
        \addlegendentry{\cite{Alberge} - $P_{\textrm{e}_2}$ ($\lambda=0.5$)}

        \addplot[blue, mark=o] 
   	table [x={x}, y={y}] {./Figs/n2_k2/bler1_SIC.txt};
   	\addlegendentry{\cite{Alberge} + \textit{SICNet} - $P_{\textrm{e}_1}$ ($\lambda=0.5$)}
        
        \addplot[dashed,blue, mark=o, mark options={solid}] 
   	table [x={x}, y={y}] {./Figs/n2_k2/bler2_SIC.txt};
        \addlegendentry{\cite{Alberge} + \textit{SICNet} - $P_{\textrm{e}_2}$ ($\lambda=0.5$)}

        \addplot[green, mark=triangle] 
   	table [x={x}, y={y}] {./Figs/n2_k2/bler1_055.txt};
   	\addlegendentry{\cite{Alberge} + \textit{SICNet} - $P_{\textrm{e}_1}$ ($\lambda=0.55$)}
        
        \addplot[dashed,green, mark=triangle, mark options={solid}] 
   	table [x={x}, y={y}] {./Figs/n2_k2/bler2_055.txt};
        \addlegendentry{\cite{Alberge} + \textit{SICNet} - $P_{\textrm{e}_2}$ ($\lambda=0.55$)}

        \addplot[cyan, mark=x] 
   	table [x={x}, y={y}] {./Figs/n2_k2/bler1_06.txt};
   	\addlegendentry{\cite{Alberge} + \textit{SICNet} - $P_{\textrm{e}_1}$ ($\lambda=0.6$)}
        
        \addplot[dashed,cyan, mark=x, mark options={solid}] 
   	table [x={x}, y={y}] {./Figs/n2_k2/bler2_06.txt};
        \addlegendentry{\cite{Alberge} + \textit{SICNet} - $P_{\textrm{e}_2}$ ($\lambda=0.6$)}
   	
 	\end{semilogyaxis}
	\end{tikzpicture}
	\caption{Different architectures $(P_\mathrm{e_{1}},P_\mathrm{e_{2}})$ versus $SNR$ performance comparison  for $\lambda=\{0.5, 0.55, 0.6\}$ in  NOMA downlink transmission with two users ($k_1=2$, $k_2=2$, $n=2$, $\Delta SNR=9$ dB).}
	\label{Fig_3}
\end{figure} 

\subsection{Comparison with the State-of-the-Art}

\subsubsection{Case 1 - $(k,n)=(2,2)$}  In order to examine performances of the proposed approach, the two-user scenario and corresponding architecture form \cite{Alberge} is recreated.
More precisely, BS jointly encodes $k_1=k_2=k=2$ bits per user ($M_c=2^{2k}=16$) and sends it over $n=2$ channel uses. 
    
    In Fig. \ref{Fig_3} we compare performances obtained with the proposed approaches and architecture replicated from \cite{Alberge}. Channel conditions for the second user can be obtained as $SNR_2=SNR_1-\Delta SNR$, where $SNR_1$ is presented on the Fig. \ref{Fig_3} $x$-axis. Significant performance improvement in terms of error performance of both users can be observed with the introduction of the \textit{SICNet} \cite{Van Luong}, compared to the baseline results in \cite{Alberge}. 
    Moreover, use of compound loss function (Eq. \ref{loss_weight}) by incorporating non-uniform users weights in the loss function (parameter $\lambda$) provides for a flexible system design, where a desirable trade-off between the two users' performance can be easily controlled through manipulation of $\lambda$ value.

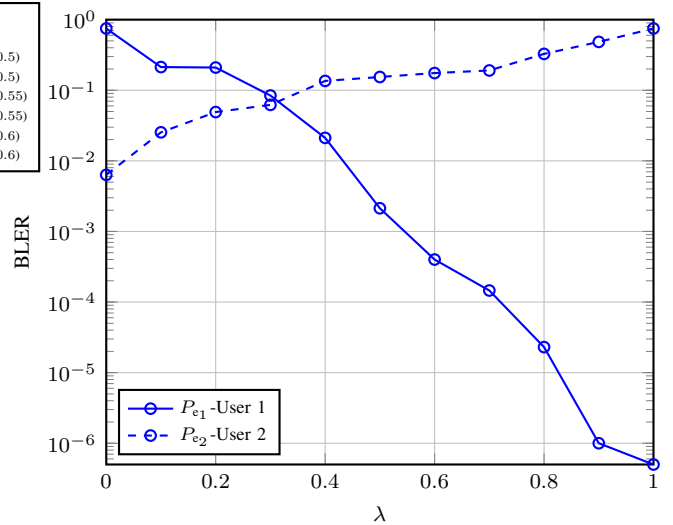
\begin{figure}[t]
\centering
\begin{tikzpicture}
  	\begin{semilogyaxis}[width=1\columnwidth, height=7.5cm, 
	legend style={at={(0.17,0.17)}, anchor= north,font=\scriptsize, legend style={nodes={scale=0.99, transform shape}}},
   	legend cell align={left},
	legend columns=1,   	 
   	x tick label style={/pgf/number format/.cd,
   	set thousands separator={},fixed},
   	y tick label style={/pgf/number format/.cd,fixed, precision=2, /tikz/.cd},
   	xlabel={$\lambda$},
   	ylabel={BLER},
   	label style={font=\footnotesize},
   	grid=major,   	
   	xmin = 0, xmax = 1,
   	ymin=0.0000005, ymax=1,
   	line width=0.8pt,
   	tick label style={font=\footnotesize},]
   	\addplot[blue, mark=o] 
   	table [x={x}, y={y}] {./Figs/lambda_bler/bler1.txt};
   	\addlegendentry{$P_{\textrm{e}_{1}}$-User 1}
   	
   	\addplot[dashed,blue, mark=o, mark options={solid}] 
   	table [x={x}, y={y}] {./Figs/lambda_bler/bler2.txt};
   	\addlegendentry{$P_{\textrm{e}_{2}}$-User 2}
 	\end{semilogyaxis}
	\end{tikzpicture}
	\vspace*{-4mm}
\caption{ $\lambda$ influence on ($P_{\textrm{e}_{1}}$, $P_{\textrm{e}_{2}}$) in two users downlink NOMA transmission ($k_1=2$, $k_2=2$, $n=2$, $SNR_1=12$ dB, $\Delta SNR=9$ dB).}
\label{Fig_lambda}
\end{figure}

     Influence of different $\lambda$ values, introduced during the training process, on system performances is illustrated in Fig. \ref{Fig_lambda}, where we plot ($P_{\textrm{e}_1}, P_{\textrm{e}_2}$) against $\mathbf{\lambda}=\{0, 0.1, 0.2, \ldots, 0.9, 1\}$. As the value of $\lambda$ increases, the loss function (Eq. \ref{loss_weight}) begins to favour the first user. This is reflected in a graceful improvement of $P_{\textrm{e}_1}$ on Fig. \ref{Fig_lambda}, simultaneously with the graceful degradation of $P_{\textrm{e}_2}$. By tuning $\lambda$, the system can be adapted to a different requirements and variable channel conditions ($\Delta SNR$). The testing phase is done on $SNR_1=12$ dB and $\Delta SNR=9$ dB, i.e., although the second user is tested on $SNR_2=3$ dB and its slope is gentler, it still outperforms the first user for $\lambda$ values below 0.3 (Fig. \ref{Fig_lambda}). 
    
\begin{figure}[t]
	\begin{tikzpicture}
  	\begin{axis}[width=1\columnwidth, height=7cm, 
	legend style={at={(0.87,0.98)}, anchor= north,font=\scriptsize, legend style={nodes={scale=0.99, transform shape}}},
   	legend cell align={left},
	legend columns=1,   	 
   	x tick label style={/pgf/number format/.cd,fixed,
   	 precision=1, /tikz/.cd},
   	y tick label style={/pgf/number format/.cd,fixed, precision=1, /tikz/.cd},
   	xlabel={In--phase},
   	ylabel={Quadrature},
   	label style={font=\footnotesize},
   	grid=major,   	
   	xmin =-2, xmax = 2,
   	ymin=-2, ymax=2,
   	line width=0.85pt,
   	tick label style={font=\footnotesize},] 
   	\addplot[red, only marks, mark=*] 
   	table [x={x}, y={y}] {./Figs/const/0.5.txt}; 
   	\addlegendentry{$\lambda=0.5$}
        tick label style={font=\footnotesize},] 
        \addplot[blue, only marks, mark=square] 
   	table [x={x}, y={y}] {./Figs/const/06.txt}; 
   	\addlegendentry{$\lambda=0.6$}
        \draw[stealth-stealth] (50,117) -- (117,50);
        \draw[stealth-stealth] (49,295) -- (117,363);
 	\node at (34,70) {User 2};
        \node at (38,345) {User 1};
  	\end{axis}
	\end{tikzpicture}
	\caption{$\lambda$ influence on learned constellations for two users NOMA downlink transmission ($k_1=2$, $k_2=2$, $n=2$, $M_c=2^{k_1+k_2}=16$).}
	\label{Fig_new}
\end{figure}
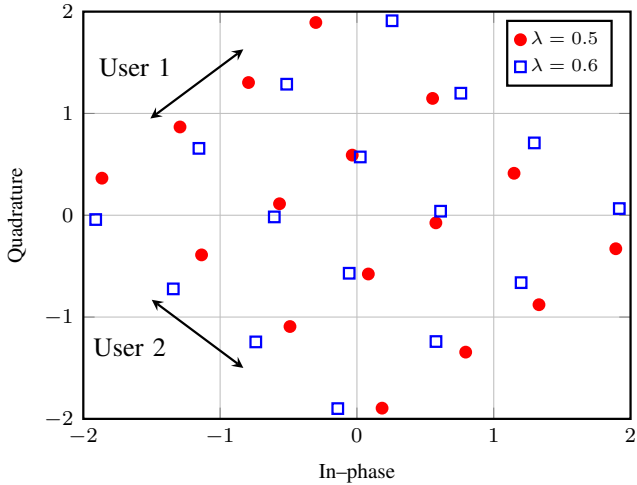

    Learned messages $\mathbf{x}$ are two-dimensional real values ($\mathbf{x}\in\mathbb{R}^2$), thus they can be visualized as symbols (or points) in 2D "constellation" (Fig. \ref{Fig_new}). 
    Fig. \ref{Fig_new} illustrates the influence of $\lambda$ on learned constellation. We observe that the distance between messages associated to user 1 increases with increasing $\lambda$ (loss function favors the first user), while the opposite happens to the second user (shape of the constellation transforms from the square ($\lambda=0.5$) to a rectangle ($\lambda=0.6$)).

\subsubsection{Case 2 - $(k,n)=(4,4)$} 
\begin{figure}[t]
	\begin{tikzpicture}
  	\begin{semilogyaxis}[width=1\columnwidth, height=7.5cm, 
	legend style={at={(0.22,0.39)}, anchor= north,font=\scriptsize, legend style={nodes={scale=0.72, transform shape}}},
   	legend cell align={left},
   	x tick label style={/pgf/number format/.cd,
   	set thousands separator={},fixed},
   	y tick label style={/pgf/number format/.cd,fixed, precision=2, /tikz/.cd},
   	xlabel={$SNR_1$},
   	ylabel={$P_{\textrm{e}}$},
   	label style={font=\footnotesize},
   	grid=major,   	
   	xmin = 10, xmax = 16,
   	ymin=0.000001, ymax=1,
   	line width=0.8pt,
   	tick label style={font=\footnotesize},]
   	\addplot[red, mark=square] 
   	table [x={x}, y={y}] {./Figs/n4_k4/bler1_fr.txt};
   	\addlegendentry{\cite{Alberge} - $P_{\textrm{e}_1}$         ($\lambda=0.5$)}
        
        \addplot[dashed,red, mark=square, mark options={solid}] 
   	table [x={x}, y={y}] {./Figs/n4_k4/bler2_fr.txt};
        \addlegendentry{\cite{Alberge} - $P_{\textrm{e}_2}$ ($\lambda=0.5$)}

        \addplot[blue, mark=o] 
   	table [x={x}, y={y}] {./Figs/n4_k4/bler1_SIC.txt};
   	\addlegendentry{\cite{Alberge} + \textit{SICNet} - $P_{\textrm{e}_1}$ ($\lambda=0.5$)}
        
        \addplot[dashed,blue, mark=o, mark options={solid}] 
   	table [x={x}, y={y}] {./Figs/n4_k4/bler2_SIC.txt};
        \addlegendentry{\cite{Alberge} + \textit{SICNet} - $P_{\textrm{e}_2}$ ($\lambda=0.5$)}

        \addplot[green, mark=triangle] 
   	table [x={x}, y={y}] {./Figs/n4_k4/bler1_055.txt};
   	\addlegendentry{\cite{Alberge} + \textit{SICNet} - $P_{\textrm{e}_1}$ ($\lambda=0.55$)}
        
        \addplot[dashed,green, mark=triangle, mark options={solid}] 
   	table [x={x}, y={y}] {./Figs/n4_k4/bler2_055.txt};
        \addlegendentry{\cite{Alberge} + \textit{SICNet} - $P_{\textrm{e}_2}$ ($\lambda=0.55$)}

        \addplot[cyan, mark=x] 
   	table [x={x}, y={y}] {./Figs/n4_k4/bler1_06.txt};
   	\addlegendentry{\cite{Alberge} + \textit{SICNet} - $P_{\textrm{e}_1}$ ($\lambda=0.6$)}
        
        \addplot[dashed,cyan, mark=x, mark options={solid}] 
   	table [x={x}, y={y}] {./Figs/n4_k4/bler2_06.txt};
        \addlegendentry{\cite{Alberge} + \textit{SICNet} - $P_{\textrm{e}_2}$ ($\lambda=0.6$)}
   	
 	\end{semilogyaxis}
	\end{tikzpicture}
	\caption{Different architectures $(P_\mathrm{e_{1}},P_\mathrm{e_{2}})$ versus $SNR$ performance comparison  for $\lambda=\{0.5, 0.55, 0.6\}$ in  NOMA downlink transmission with two users ($k_1=4$, $k_2=4$, $n=4$, $\Delta SNR=9$ dB).}
	\label{Fig_4}
\end{figure}
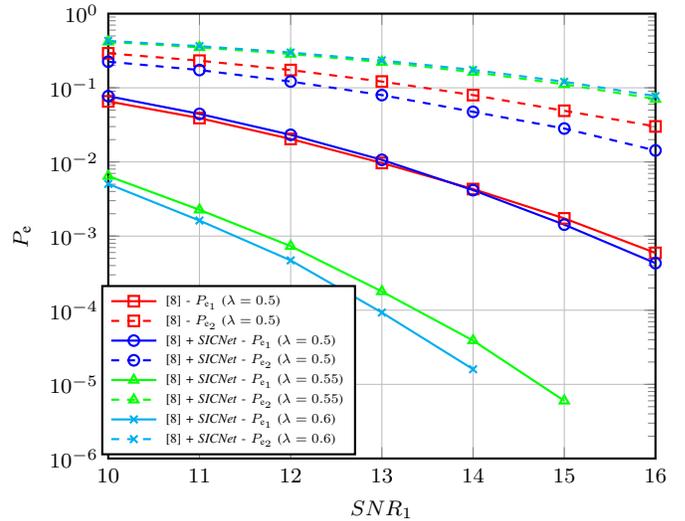

Most of the existing works targeting AE-based downlink NOMA restrict their attention to the case of two-bit user messages ($k=2$) \cite{Alberge, VanLuong_2}. Exploring how learned NOMA will behave in higher dimensions, we further expand the number of bits per user to $k_1=k_2=k=4$ ($M_c=2^{2k}=256$) and the number of channel uses to $n=4$. 
Although the number of messages grows exponentially, the proposed approach significantly outperforms an architecture from \cite{Alberge} (Fig. \ref{Fig_4}).
More precisely, from Fig. \ref{Fig_4} we observe that \textit{SICNet} receiver noticeably improves the performance of the second user (when $\Delta SNR$ is known), while the first user performance remains comparable to \cite{Alberge}.
On the other hand, with the introduction of the compound loss function ($\lambda$, Eq. \ref{loss_weight}), the proposed approach significantly outperforms \cite{Alberge} in terms of the first user performance (Fig. \ref{Fig_4}, green and cyan solid lines), with the price of slight performance degradation for the second user.
    
We note that moving to higher constellation dimensions (i.e., from $n=2$ to $n=4$), although incurs higher complexity ($M_c=256$), leads to a considerable improvement of error performance of both users. It also leads to faster saturation of the second user performance with the increase of $\lambda$ (i.e., already for $\lambda>0.55$ on Fig. \ref{Fig_4}).
    
\section{Conclusion}

In this paper, we presented a flexible and efficient weighted AE-based method for design of downlink NOMA constellations. The method demonstrates promising performance under simplicity in training and tuning to the desired error probability balance between users. In the future work, we plan explore a combination of outer low-density parity-check codes (LDPC) with inner weighted AE-based dowlink NOMA constellations.

\end{document}